\documentclass{aa}
\usepackage{graphicx}
\usepackage{txfonts}

\newcommand{\xte}{{\textit{RXTE}}}

\newcommand{\swift}{{\textit{Swift}}}
\newcommand{\msun}{{\rm M}_{\sun}}

\newbox\grsign \setbox\grsign=\hbox{$>$} \newdimen\grdimen \grdimen=\ht\grsign
\newbox\simlessbox \newbox\simgreatbox \newbox\simpropbox
\setbox\simgreatbox=\hbox{\raise.5ex\hbox{$>$}\llap
     {\lower.5ex\hbox{$\sim$}}}\ht1=\grdimen\dp1=0pt
\setbox\simlessbox=\hbox{\raise.5ex\hbox{$<$}\llap
     {\lower.5ex\hbox{$\sim$}}}\ht2=\grdimen\dp2=0pt
\setbox\simpropbox=\hbox{\raise.5ex\hbox{$\propto$}\llap
     {\lower.5ex\hbox{$\sim$}}}\ht2=\grdimen\dp2=0pt
\def\ga{\mathrel{\copy\simgreatbox}}
\def\la{\mathrel{\copy\simlessbox}}

\begin{document} 

   \title{Long-term quasi-periodicity of 4U~1636--536 resulting from accretion disc instability}

   
   \author{Mateusz  Wi\'sniewicz\inst{\ref{inst1}} \and Agnieszka S\l{}owikowska\inst{\ref{inst1}} \and Dorota Gondek-Rosi\'nska\inst{\ref{inst1}} \and Andrzej A. Zdziarski\inst{\ref{inst2}} \and\\ Agnieszka Janiuk\inst{\ref{inst3}}}
   
   \institute{Institute of Astronomy, University of Zielona G\'ora, Szafrana 2, PL-65-516 Zielona G\'ora, Poland \\ \email{mateusz@astro.ia.uz.zgora.pl}\label{inst1}
   \and Centrum Astronomiczne im.\ M. Kopernika, Bartycka 18, PL-00-716 Warszawa, Poland\label{inst2}
   \and Centre for Theoretical Physics, Polish Academy of Sciences, Al.\ Lotnik\'ow 32/46, PL-02-668 Warsaw, Poland\label{inst3}}
   
  \abstract{We present the results of a study of the low-mass X-ray binary 4U~1636--536. We have performed temporal analysis of all available \xte/ASM, \swift/BAT and MAXI data. We have confirmed the previously discovered quasi-periodicity of $\simeq\! 45$ d present during $\sim$2004, however we found it continued to 2006. At other epochs, the quasi-periodicity is only transient, and the quasi-period, if present, drifts. We have then applied a time-dependent accretion disc model to the interval with the significant X-ray quasi-periodicity. For our best model, the period and the amplitude of the theoretical light curve agree well with that observed. The modelled quasi-periodicity is due to the hydrogen thermal-ionization instability occurring in outer regions of the accretion disc. The model parameters are the average mass accretion rate (estimated from the light curves), and the accretion disc viscosity parameters, $\alpha$, for the hot and cold phases. Our best model gives relatively low values of $\alpha_{\rm cold}\simeq 0.01$ and $\alpha_{\rm hot}\simeq 0.03$. }

   \maketitle
%
\section{Introduction}

4U~1636--536 is a low-mass X-ray binary (LMXB) discovered by \citet{will1974}. The photometry of the optical counterpart (V801~Ara) shows a short orbital period of 3.79 h \citep{giles02}. The binary system consists of a late-type, low-mass ($\simeq 0.3$--$0.4\,\msun$) donor, which transfers mass onto a neutron star \citep{fuj1986,van1990}. \citet{gall2006} estimated the distance to 4U~1636--536 to be $D=6.0 \pm 0.5\,{\rm kpc}$ from Eddington limited X-ray bursts, assuming the neutron star mass of $1.4\,\msun$ and the stellar radius of $10$~km.  According to \citet{cas2006}, the mass function and mass ratio of 4U~1636--536 are $f(M) = 0.76 \pm 0.47\,\msun$ and $M_2/M_{\rm NS} \simeq 0.21$--0.34, respectively, where $M_{\rm NS}$ is the mass of the neutron star and $M_2$ is the mass of the donor. They also estimated the inclination as $i\simeq 36\degr$--$60\degr$. The binary is a persistent X-ray source, although it shows significant flux variations on both long and short time scales. On time scales of hours, its flux varies by a factor of $\sim$2--3 \citep{hoff1977,oha1982,bre1986,has1989}. The presence of kHz quasi-periodic oscillations, which are also visible in the system during X-ray bursts, shows that the neutron star has been spun-up through accretion \citep{zha1996,stro1999}.

\begin{figure}
\centerline{\includegraphics[width=90mm]{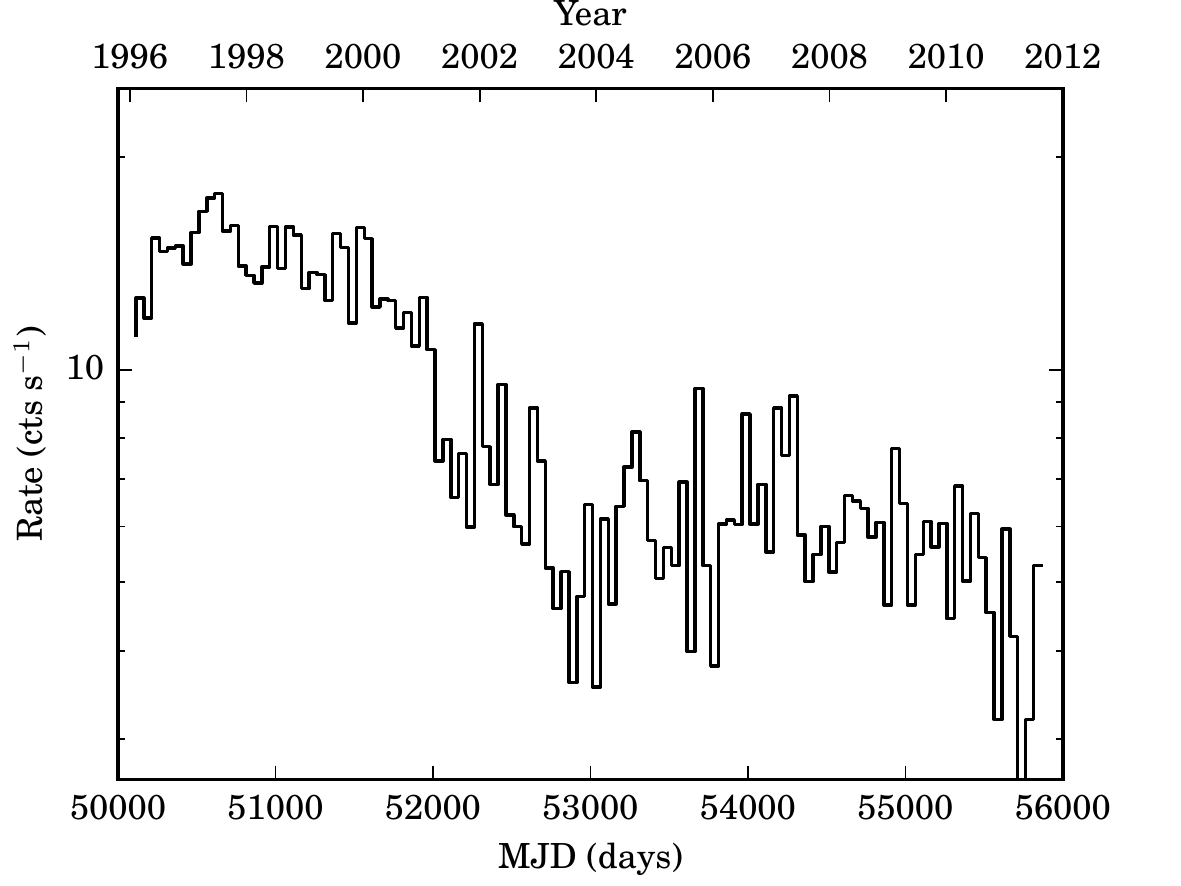}}
\caption{The 50-d average \xte/ASM light curve of 4U~1636--536 in the energy range of 1.3--12.2\,keV from January 1996 to November 2011.}
\label{50day}
\end{figure}

\begin{figure*}
\centerline{\includegraphics[width=140mm]{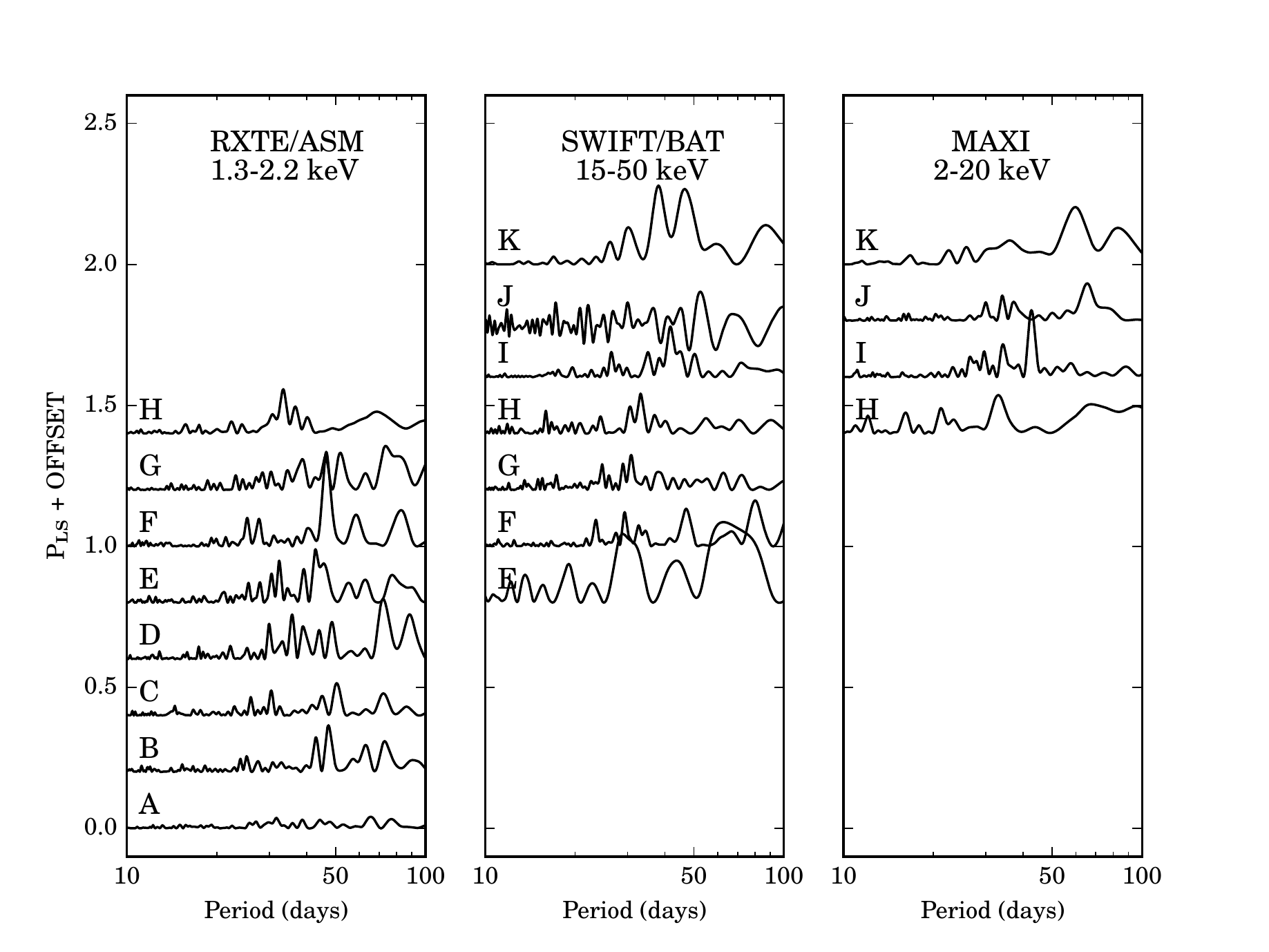}}
\caption{The power spectra (with offsets) for the \xte/ASM (left panel), the \swift/BAT (middle panel) and the MAXI (right panel) in the time intervals given in MJD by: (A) 50088--50741, (B) 50742--51452, (C) 51453--52162, (D) 52163--52886, (E) 52887--53564, (F) 53565--54185, (G) 54186--54749, (H) 54750--55322, (I) 55323--55869, (J) 55893--56461, (K) 56462--56738.}
\label{all_periodograms}
\end{figure*}

4U~1636--536 has been monitored daily in the 1.3--12.2 keV energy range by the All Sky Monitor (ASM) on-board of the {\it Rossi X-ray Timing Explorer\/} (\xte) from 1996 until 2011. During the first four years of \xte/ASM observations (1996--2000) the source count rate was relatively stable at $\sim 15~{\rm cts~s^{-1}}$. After 2000, it started to gradually decline and occasionally show a statistically significant quasi-periodic variability \citep{shih2005}. Those authors reported the presence of a long-period, $\simeq 47~{\rm d}$, quasi-periodic variability in the 2004 light curve. They suggested that the observed flux variability is caused by the variability of the accretion flow related to X-ray irradiation of the disc.

In our paper, we study the variability of 4U~1636--536 taking into account the currently available data from three X-ray monitors, spanning almost 20 years (1996--2014). We interpret the data in terms of a disc instability model. In Sect.\ \ref{analysis}, we describe the X-ray data and perform their timing analysis. In Sect.\ \ref{theory}, we present the theoretical model of evolution of an accretion disc around a neutron star used in the paper. We model the evolution of the accretion disc for a wide range of viscosity parameters.  We find that the observed quasi-periodicity can be explained by the thermal instability due to the ionisation of hydrogen. We discuss our results, and finally, in Sect.\ \ref{conclusions}, we give our main conclusions.

\section{Data analysis}
\label{analysis}

We analyse the X-ray data obtained with the \xte/ASM \citep{brs93,levine96}, the Burst Alert Telescope (BAT) on board of \swift\/ \citep{m05,krimm_2013} and the Monitor of All-sky X-ray Image (MAXI) on board of {\it International Space Station\/} \citep{maxi_info}. One-day average data from these instruments have been used to analyse the flux variability at different epochs. The ASM operated in the energy range of 1.3--12.2\,keV with the sensitivity of $\sim 30$\,mCrab. The BAT observes $88\%$ of the sky each day with the sensitivity of $\sim 5$\,mCrab in the energy range of 15--50\,keV. The \swift/BAT has observed 4U~1636--536 since 2005. The Japanese X-ray camera MAXI has observed X-ray sources since August 2009, scanning most of the sky every 96 minutes. It operates in the energy range of 2--20\,keV with the sensitivity of $20~{\rm mCrab}$.

The X-ray flux history in the ASM data is shown in Fig.~\ref{50day}. The light curve can be divided into three main parts. From 1996 to 2000, the flux was relatively stable ($\sim 15~{\rm cts~s^{-1}}$), with only some low-amplitude variability. Between 2000 and 2002, the flux gradually declined to a level below $\sim 10~{\rm cts~s^{-1}}$. Since 2003, the flux was below $\sim 10~{\rm cts~s^{-1}}$ and high-amplitude variability was present. 

We have linearly detrended all the data and then applied a Fourier transform. We use the standard Lomb-Scargle method \citep{sca1982}, as implemented in the \textsc{astroML} \citep{astroml} Python library. This method is suitable for our unevenly sampled time series. We have divided all the data into 11 intervals with the length of $\simeq 2$\,yr each, denoting them with the capital letters: A: 1996--1997, B: 1997--1999, C: 1999--2001, D: 2001--2003, E: 2003--2005, F: 2005--2007, G: 2007--2008, H: 2008--2010, I: 2010-2011, J: 2011-2013, and K: 2013-2014. The power spectra of these intervals for the data obtained with three instruments are shown in Fig.\ \ref{all_periodograms}.

\begin{figure}
\centerline{\includegraphics[width=85mm]{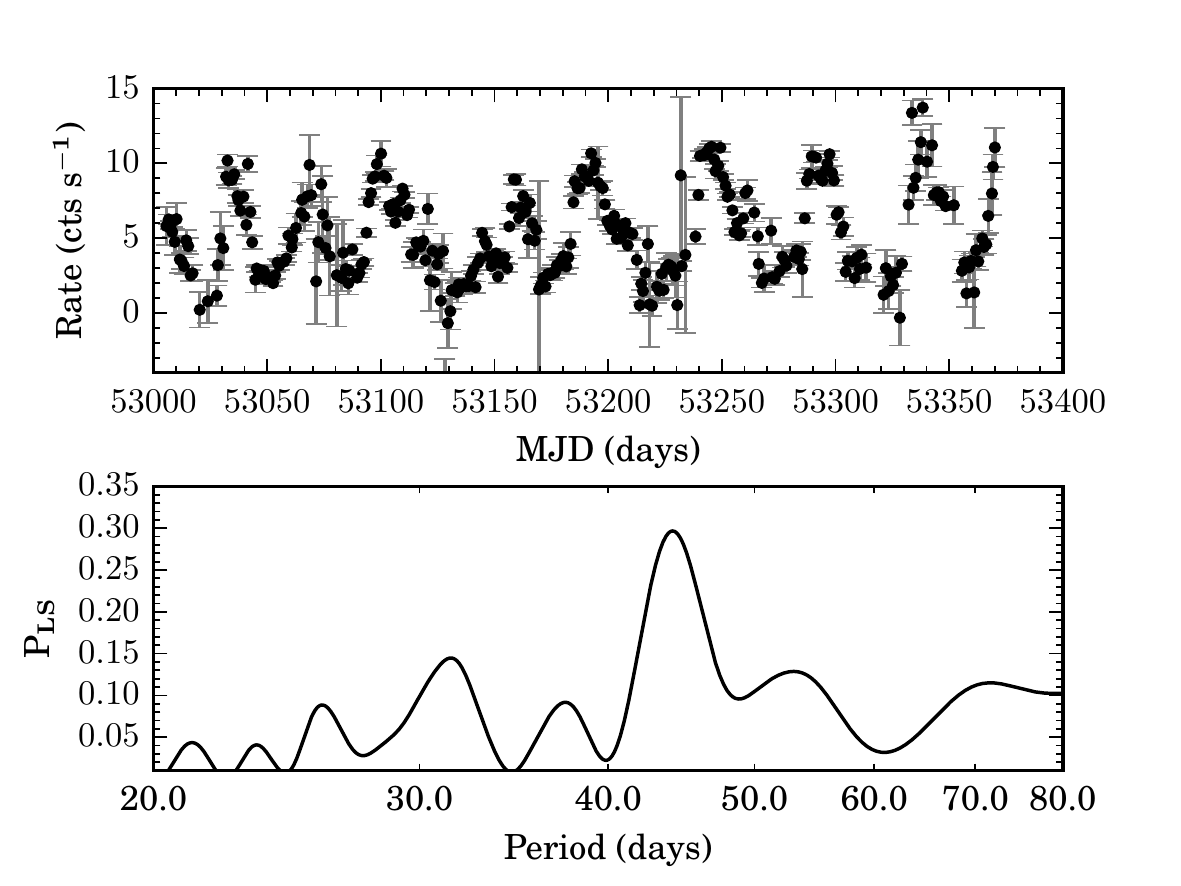}}
\caption{Top: one-day average \xte/ASM light curve for 2004. 
Bottom: the corresponding power spectrum. }
\label{2004}
\end{figure}
\begin{figure}

\centerline{\includegraphics[width=85mm]{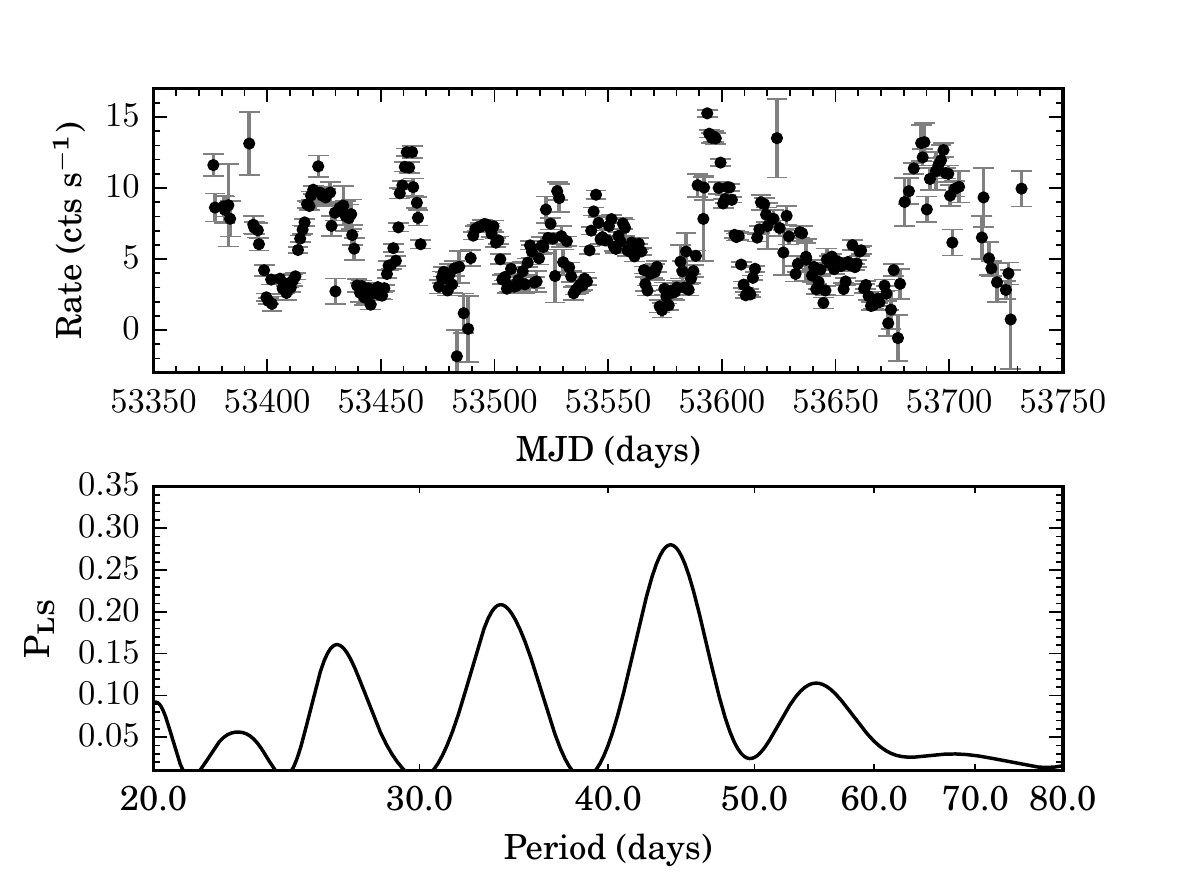}}
\caption{The same as in Fig.\ \ref{2004} but for 2005.}
\label{2005}
\end{figure}
\begin{figure}
\includegraphics[width=85mm]{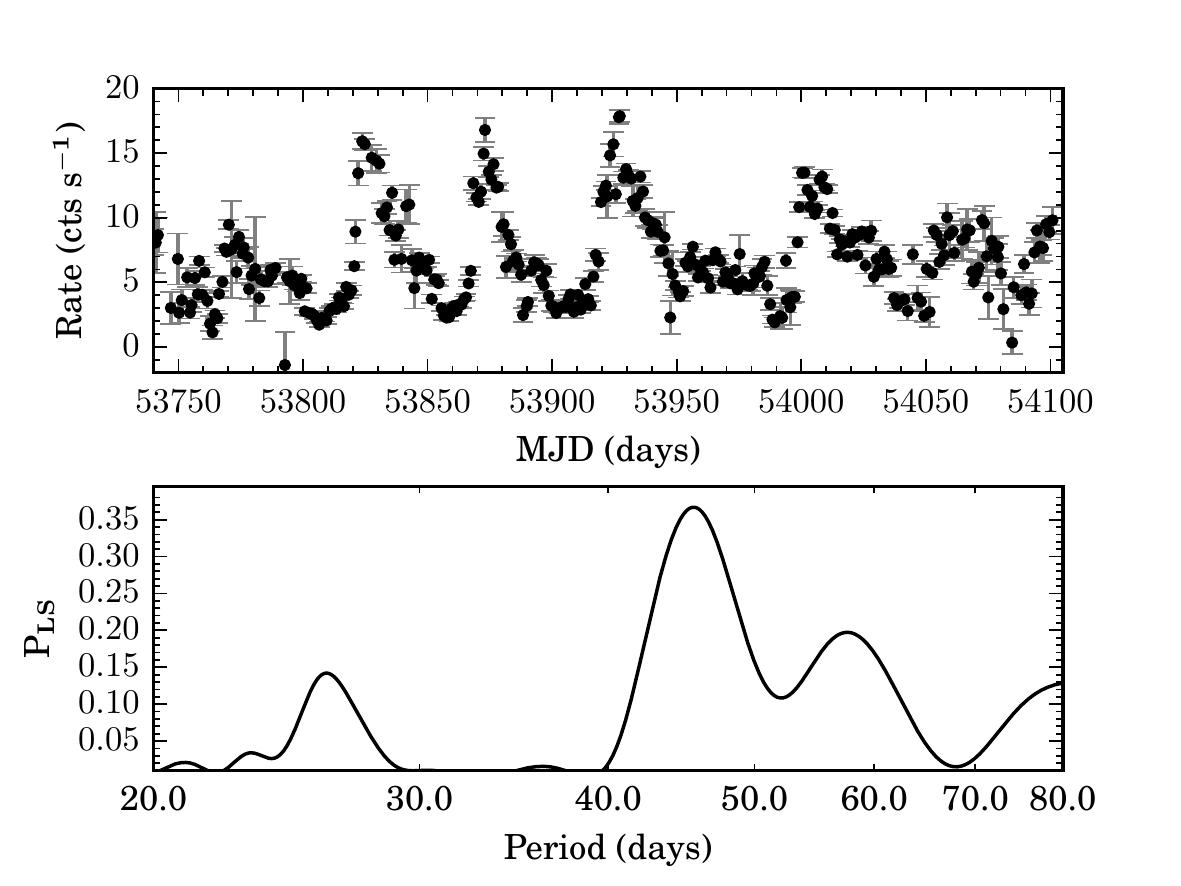}
\caption{The same as in Fig.\ \ref{2004} but for 2006.}
\label{2006}
\end{figure}

\begin{figure}
\centerline{\includegraphics[width=85mm]{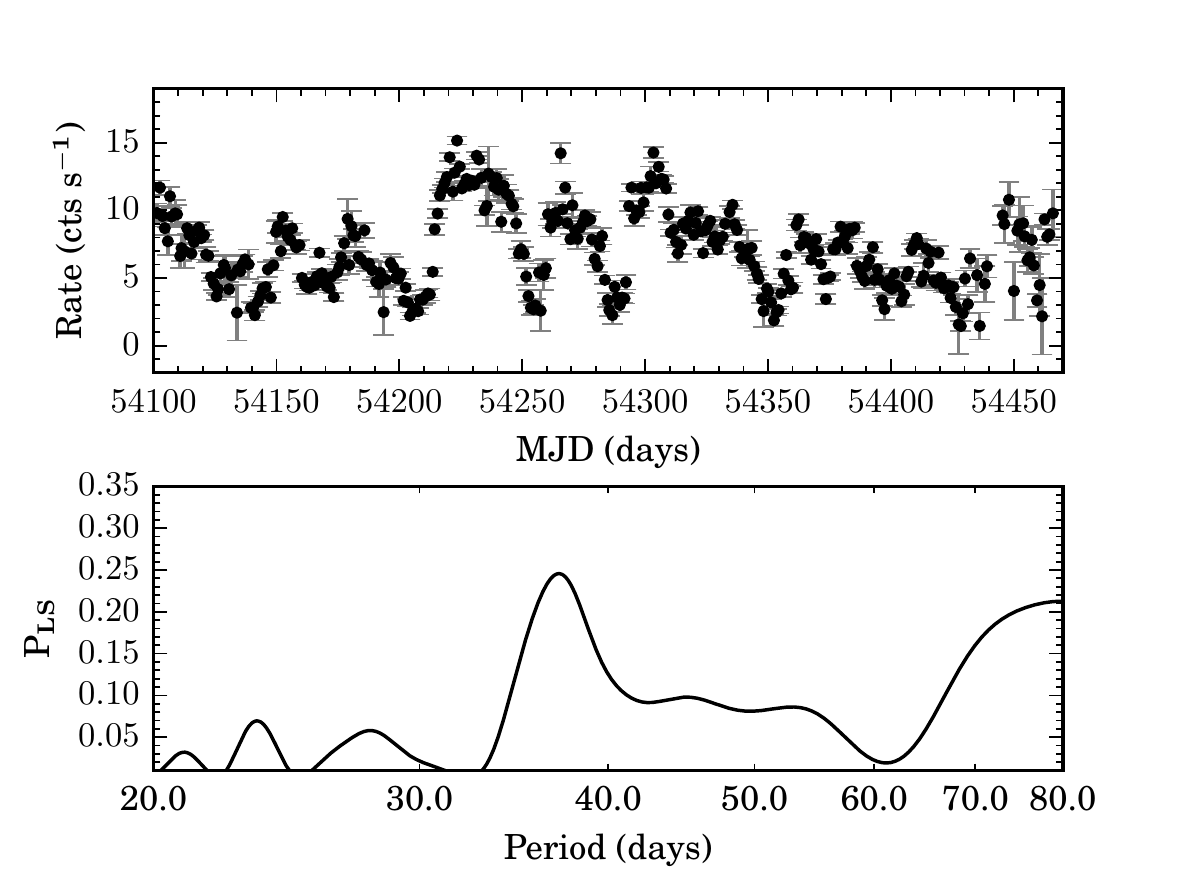}}
\caption{The same as in Fig.\ \ref{2004} but for 2007.}
\label{2007}
\end{figure}

\begin{figure}
\centerline{\includegraphics[width=90mm]{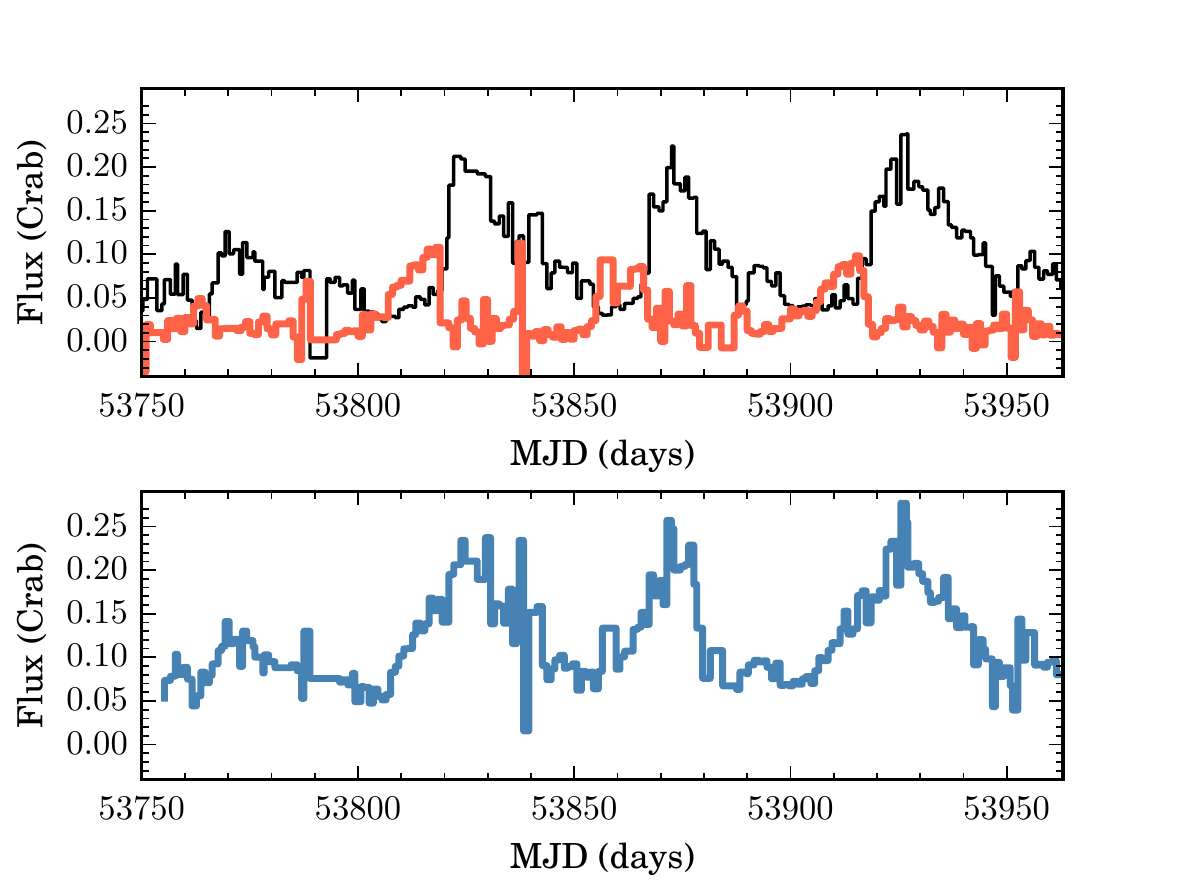}}
\caption{Top: one-day average ASM 1.3--12.2\,$\rm keV$ (thin black curve) and
BAT 15--50~$\rm keV$ (thick red line) light curve of 4U~1636--536
from MJD 53755 to 53963 in Crab units. Bottom: sum of one-day average ASM and BAT light curves (1.3--50\,$\rm keV$). }
\label{flux_asm_bat}
\end{figure}

We then focus on 2-yr intervals from the ASM data shown in the Fig.\ \ref{all_periodograms}. We see no periodicity in the interval A. From B to D, some quasi-periodicity appears, but is not statistically significant, with the peaks in the power spectra being low. Finally, from the interval E (2003--2005), significant quasi-periodicity is visible.  A similar periodicity was noticed by \cite{shih2005} for the 2004 epoch. The quasi-periodic variability began to gradually decline from the interval G.  

In order to perform a more detailed analysis, for the time span where significant quasi-periodicity is visible, we focus on the power spectra of the ASM data from the second half of the interval E to the first half of the interval G, i.e. 2004--2007. Therefore, for this time span the data were divided into 1-yr intervals. The light curves and the corresponding power spectra are shown in Figs.\ \ref{2004}--\ref{2007}. Relatively strong quasi-periodic variability is visible in 2004, 2005 and 2006, where the power spectra maxima are at $P = 44.5~{\rm d}$, 44.0 d and 45.5 d, respectively. In 2007, the Lomb-Scargle power significantly decreased to $P_{\rm LS}\sim 0.25$ but also the power-spectrum maximum shifted to around 37 days, see Fig.\ \ref{2007}.

The BAT data are relatively noisy. We see some quasi-periodicity with $P \simeq 45~{\rm d}$ in the interval F. Then, it seems that the period of the variability is decreasing until the interval I, where the 45-d periodicity is again visible. 

In order to approximate the bolometric light curve of 4U~1636--536, the fluxes of the ASM and BAT in the Crab unit were summed. We show the individual light curves for soft and hard X-ray ranges and their sum in Fig.\ \ref{flux_asm_bat} for the time span of MJD~53755--53963 within the interval F. These are the most predominant periodic data in our whole sample and therefore this part of the light curve was later used as a template for our theoretical model.  The apparent delay of the ASM light curve with respect to that of the BAT of about 10 d (calculated by cross-correlation) appears to be related to an anti-correlation between these bands reflecting the spectral softening from the hard spectral state to the soft one with the pivot energy of $\sim 10$ keV, see e.g. \citet{done2007} and references therein. In an ideal case, the maximum of the flux above the pivot would be at the minimum flux below the pivot, which corresponds to the delay of a half of the period. In our case the pivot is within the ASM energy range, which then reduces the delay.  

We see some $\sim$40-d quasi-periodicity in the first 2-yr interval, H, of the MAXI data. In the next interval, I, there is a strong quasi-periodicity with $P \simeq 45~{\rm d}$, similarly to the corresponding \swift/BAT data. Later, this periodicity disappears and another one with approximately doubled period, i.e. $\simeq$70\,d, appears.

\section{The theoretical model}
\label{theory}

The theory of geometrically thin, stationary, accretion disc of \citet{shakura1973} describes well accretion as the source of energy of many astrophysical objects. On the other hand, many of the observed accretion sources are strongly variable, which may be related to non-stationary effects predicted theoretically. In particular, \citet{meyer1982} and \citet{smak1982} independently found that accretion discs become unstable in the range of temperatures corresponding to ionization of hydrogen. This instability is the result of an inverse relation between the temperature and opacity in the range between $\sim\! 10^3$ K and $\sim\! 10^4$ K. The Rosseland mean opacity is smoothly decreasing with increasing temperature at higher temperatures. But when the temperature drops below $10^4$ K, hydrogen partially recombines, and the opacity steeply decreases with decrease of temperature. Then, at $T \la 10^3$ K, hydrogen becomes completely neutral, and the huge grain opacity dominates \citep{alexander1975}.

The theory of partial hydrogen ionisation instability was originally developed to explain the large-amplitude luminosity variations of cataclysmic variables (dwarf novae; \citealt{meyer1982,smak1984}). It is also believed that the same mechanism is responsible for outbursts in soft X-ray transients \citep{cannizzo1982,dubus2001,lasota2001,baginska2014} and active galactic nuclei \citep{lin1986,clarke1988,mineshige1990,siemiginowska1996}.

Also, \citet{hameury98} studied time dependent numerical models of the accretion disc evolution to explain the cyclic outbursts of dwarf novae and soft X-ray transients. They used an adaptive grid technique, in which the outer radius of the disc, $R_{\rm out}$, is not fixed but it varies with time. In their simulations, the variable $R_{\rm out}$ allows for a propagation of the heat front from outer disc radii inwards, which is aimed to explain the properties of the dwarf nova SS Cygni.

Similarly, \citet{dubus2001} used the adaptive grid numerical scheme. A large number of grid points, at least about 100--800, is required in such computations in order to conserve the mass of the disc, which was checked a posteriori by integration of the surface density over radius. In our computations, the outer disc radius is fixed, and the mass is conserved. We use a linear grid in $2\sqrt{r}$. We keep the number of grid points at 150, to allow for moderate resolution. In this way, we assure that the thickness of the disc is not larger than the size of the disc cell, $\Delta r$, so the assumption of the geometrically thin disc is satisfied. This condition is important because we calculate the 1+1D structure (see below) rather than perform 2D hydrodynamical simulations. Still, when we increase the number of grid points to, e.g., 300, the obtained light curve is only moderately affected, keeping the qualitative character of the variability.

\citet{ls00,ls08} presented an analytical model of time dependent accretion disc, in which the continuity equation for the surface density change was solved. In this way, the authors explained the power-law decay of the luminosity and model the light curves of the X-ray novae.  These authors do not compute numerically the thermal-viscous evolution and cyclic outbursts of the sources, because this would require solving the time-dependent thermal balance. Nevertheless, \citet{sls08} successfully explained with such a model the decay profiles of the black-hole transients A 0620--00 and GX 1124--68. In our computations, we solve both the time-dependent evolution of the disc density and temperature; thus, modeling of cyclic outbursts and quiescent phases is possible. 

We use the method of \citet{smak1984} in order to calculate the time evolution of an accretion disc around a neutron star. We follow the changes of the disc temperature and density. Then the surface temperature determines the emitted luminosity, observable in the X-ray band. We do not explicitly take into account the emission of the boundary layer, but include it in the estimate of the total luminosity as a function of the rate of mass supply at the disc outer radius.

\subsection{Vertical structure of an accretion disc}

In a geometrically thin accretion disc, the vertical and radial structures are decoupled \citep{pringle1981}. The equations that describe the disc vertical structure are very similar to those describing the internal structure of a star. 
For a Keplerian disc with the angular velocity, $\Omega = (G M/R^3)^{1/2}$, we have the energy balance,
\begin{equation}
 \frac{{\rm d}F}{{\rm d}z} = -\alpha P \frac{{\rm d} \Omega}{{\rm d}R} ,
\end{equation}
where $M$ is the mass of the accretor, $F$ is the energy flux, $\alpha$ is the viscosity parameter, $P$ is the pressure, and $z$ and $R$ are the height and radius of the accretion disc, respectively. Then, the hydrostatic equilibrium reads,
\begin{equation}
 \frac{1}{\rho} \frac{{\rm d}P}{{\rm d}z} = - \Omega^2 z\equiv g_z, 
\end{equation}
where $g_z$ is the vertical component of gravity. The energy transfer equation can be written as
\begin{equation}
\frac{{\rm d} \ln T}{{\rm d}z} = \frac{{\rm d}\ln P}{{\rm d}z}\nabla 
\end{equation}
where $T$, $P$ and $\rho$ are temperature, pressure and density, respectively, and $\nabla\equiv {\rm d} \ln T/{\rm d}\ln P$. In an optically-thick disc without convection, it is the radiative gradient,
\begin{equation}
\nabla =\nabla_{\rm rad}= \frac{\kappa P F}{4 P_{\rm rad} c g_z}
\end{equation}
where $\kappa$ is the Rosseland mean opacity and $P_{\rm rad}$ is the radiation pressure. However, in the case of partially ionized hydrogen, a major mode of energy transport may be due to convection, $\nabla_{\rm conv}$, which acts together with the radiative transport. We use here $\nabla = \nabla_{\rm conv}+\nabla_{\rm rad}$, where $\nabla_{\rm conv}$ is calculated using the mixing-length  theory, as in stellar models \citep{paczynski1969}.

The equations are integrated between the disc midplane, $z=0$, and the disc photosphere at the height $H$, given by the optical depth of the top layer of $\int_H^\infty\rho \kappa {\rm d}z=2/3$, where $\rho$ is the mass density (see \citealt{rozanska1999}).  The tables of opacities are taken from \citet{cs69,cs70} and \citet{seaton1994} for temperatures higher than $\log_{10} T > 3.5$ and from \citet{alexander1983} for lower temperatures. At the photosphere, the effective temperature of the disc is $T_{\rm eff}= T(H)$, and $\sigma T^4_{\rm eff} = Q^{+}=Q^{-}$ is resulting from the heating of the disc through the viscous energy dissipation, $Q^+$, balanced by the radiative cooling, $Q^-$, where $\sigma$ is the Stefan-Boltzmann constant.  The integration of equations for the vertical structure gives the surface density, $\Sigma$,
\begin{equation}
 \Sigma = 2 \int_0^H \rho {\rm d}z.
\end{equation}

\begin{figure}
\centerline{\includegraphics[width=90mm]{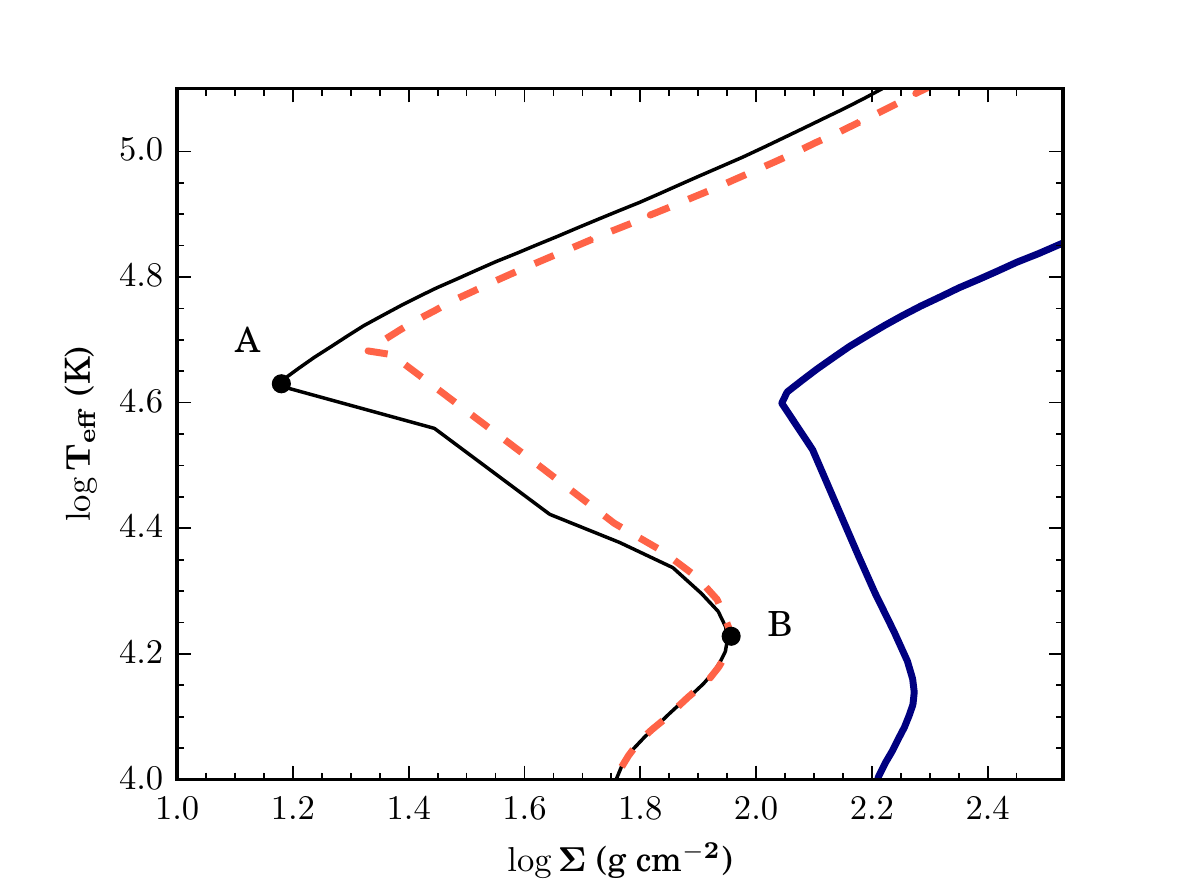}}
\caption{An example of the surface density--effective temperature $\Sigma$-$T_{{\rm eff}}$ relations (S-curves) for an accretion disc around a $1.4\,\msun$ neutron star at $R = 2\times 10^4\,R_{\rm g}$ (solid black and red dashed curves) and $2\times 10^5\,R_{{\rm g}}$ (thick blue curve). The black curve corresponds to $\alpha_{\rm cold} = 0.01$, $\alpha_{\rm hot} = 0.05$, the red dashed curve, to $\alpha_{\rm cold} = 0.01$, $\alpha_{\rm hot} = 0.03$, and the thick blue curve, to $\alpha_{\rm cold} = \alpha_{\rm hot}=0.01$. For the black curve, the hydrogen ionisation instability begins in A and ends at B; analogous points can be determined for other curves. }
\label{scurve}
\end{figure}

The viscous energy dissipation rate per unit time and volume is 
\begin{equation} 
Q^+ = \frac{3}{2} \alpha P_{{\rm tot}} \Omega , 
\end{equation}
where we assumed that the viscous stress tensor $R$-$\phi$ component is proportional to the total pressure $P_{{\rm tot}}$ (sum of the gas and radiation pressures), and scaling with $\alpha$  \citep{shakura1973}. 
Since we know the relation between the effective and central temperatures from the energy transfer equation, we can represent the energy balance in a $\Sigma$-$T_{{\rm eff}}$ diagram. The $T_{{\rm eff}}$ is directly related to the accretion rate, $\dot{M}$, which average is given by external conditions (i.e., the mass transfer rate from the companion star) and which is a parameter of the model. Example solutions are plotted in Fig.\ \ref{scurve} for the accretor mass of $M_{\rm NS}=1.4~\msun$ and selected values of the viscosity parameter, $\alpha$, and the radius, $R$.

We can distinguish three characteristic regions in the S-curve. The lower branch corresponds to thermally stable and cool neutral hydrogen. The upper branch corresponds to thermally stable but hot gas, where hydrogen is fully ionised. The opacity is then due to bound-free transitions in heavy elements, free-free transitions and electron scattering. Finally, the middle branch corresponds to partially-ionised, thermally unstable region of the disc \citep{bath1982,faulkner1983}.

Calculating the time-dependent radial structure of an accretion disc can be significantly simplified because of the universality of the above properties. The critical column densities, $\Sigma_{\rm min}$ and $\Sigma_{\rm max}$, between which the unstable branch of the stability curve appears (i.e. the points A and B in Fig.\ \ref{scurve}), are functions of the accretor mass (fixed in our case), the viscosity parameter, and the radius. On the other hand, the corresponding critical values of the effective temperature are given by the temperatures at which hydrogen becomes dominantly neutral or ionized, and thus are nearly universal, i.e., only weakly depend on the accretion flow parameters. 

\citet{smak1984} suggested the following simple scaling relations, assuming that the shape of the stability curve does not change and the critical points shift with the radius throughout the disc,
\begin{eqnarray}
\lefteqn{
 \log T_{{\rm eff}}^{{\rm A}} = \log T_{{\rm eff}}^{{\rm B}} + a_{\rm r} \log (R/2\times 10^4 R_{\rm g}) + a_\alpha \log (\alpha/0.1),}\\
\lefteqn{
 \log \Sigma^{{\rm A}} = \log \Sigma^{{\rm B}} + b_{\rm r} \log (R/2\times 10^4 R_{\rm g}) + b_\alpha \log (\alpha/0.1),}
\end{eqnarray}
where $2\times 10^4 R_{\rm g}$ is a chosen reference radius and $R_{\rm g}\equiv GM/c^2$ is the gravitational radius. In our procedure, we choose two pairs of values of $R$ and $\alpha$, for which we calculate the $\Sigma$-$T_{\rm eff}$ curves. We then calculate the positions of the A and B points, and then fit the coefficients to the stability curves calculated at intermediate values. The resulting values are $a_{\rm r}=-0.09$, $a_\alpha=-0.01$, $b_{\rm r}=1.11$, $b_{\rm \alpha}=-0.73$.

In order to reproduce the observed amplitudes of dwarf novae outbursts, it was suggested that the viscosity parameter $\alpha$ must have different values on the upper and lower branches of stability curve (see, e.g., \citealt{lasota2001}, \citealt{done2007}, for reviews). In our computations, we also assume it, implying $\Sigma_{\rm min}=\Sigma^{{\rm A}}\simeq \Sigma(\alpha_{\rm hot})$ and $\Sigma_{\rm max}=\Sigma^{{\rm B}}\simeq \Sigma(\alpha_{\rm cold})$. A bridging formula is used for the transition between the two values of $\alpha$ along the stability curve,
\begin{equation}
\log_{10}(\alpha)= \log_{10}(\alpha_{\mathrm{cold}}) + \frac{\log_{10}(\alpha_{\mathrm{hot}}/\alpha_{\mathrm{cold}})}{1 + \left(2.5 \times 10^4\,{\rm K}/T_\mathrm{c}\right)^8}.
\end{equation}

The above old, observationally motivated, approach, was confirmed by numerical simulations of the magneto-rotational instability (MRI, \citealt{balbus1991}). This instability is supposedly the physical process that is acting behind the viscosity mechanism in astrophysical flows. The non-linear development of the MRI leads to the disc turbulence, and hence turbulent viscosity, because magnetic fields are frozen into the fluid in a differentially rotating disc. Thus, the MRI provides an efficient way to transport the angular momentum outwards in the disc and allow for accretion. It has been shown, e.g., in \citet{janiuk2004}, that the resistive diffusion of magnetic field, as quantified by the magnetic Reynolds number, is low in a cold state of the discs of stellar mass accreting compact objects (white dwarfs, neutron stars and black holes in binaries). Therefore, it is expected that the MRI turbulence disappears in the quiescent disc. In the outburst state, on the other hand, the action of MRI turbulence is efficient. Consequently, the approach postulated for these systems, namely the $\alpha_{\rm cold} < \alpha_{\rm hot}$ is justified, and we follow it here.

In our numerical approach, we first calculate the disc vertical structure as a sequence of local solutions. We then calculate the time-dependent radial structure, which yields the final light curve. This method is described in detail by \citet{smak1984} and \citet{siemiginowska1996}.

\subsection{Results and discussion}
\label{results}

Following \cite{gall2006}, we assume $M_{\rm NS} = 1.4\,\msun$, $R_{{\rm NS}} = 10$\,km, and $D = 6.0$\,kpc. We have estimated the average bolometric flux by scaling the ASM and BAT spectra to the Crab, adding them and then applying a correction for the emission beyond the energy ranges of those instruments of 50\%. This yields $F_{{\rm bol}} \simeq 4.4 \times 10^{-9}$ erg cm$^{-2}$ s$^{-1}$, and, assuming isotropy, $L_{\rm bol}=4\pi D^2 F_{\rm bol}\simeq 1.9\times 10^{37}$ erg s$^{-1}$. The accretion efficiency in the Newtonian approximation (e.g., \citealt{frank2002}) is
\begin{equation}
 \eta = \frac{GM_{\rm NS}}{R_{{\rm NS}}c^2}\simeq 0.21,
\end{equation}
which implies the accretion rate averaged over an outburst of
\begin{equation}
 \dot{M} = \frac{L_{\rm bol}}{\eta c^2}\simeq 1.6 \times 10^{-9}\,\msun\,{\rm yr}^{-1}.
\end{equation}
We assume $\dot M=1.5 \times 10^{-9}\,\msun\,{\rm yr}^{-1}$ in the calculations. We study models with $0.001\leq \alpha_{\rm cold}\leq 0.1$ and with the ratio of $0.14 \leq \alpha_{\rm cold} / \alpha_{\rm hot}\leq 1.0$. We assume the disc outer radius equals $R_{\rm out} = 2\times 10^5 R_{\rm g}$, which corresponds to about 80 per cent of the Roche-lobe radius of the neutron star. The disc inner radius is at $6 R_{\rm g}$.

\begin{figure}
\centerline{\includegraphics[width=85mm]{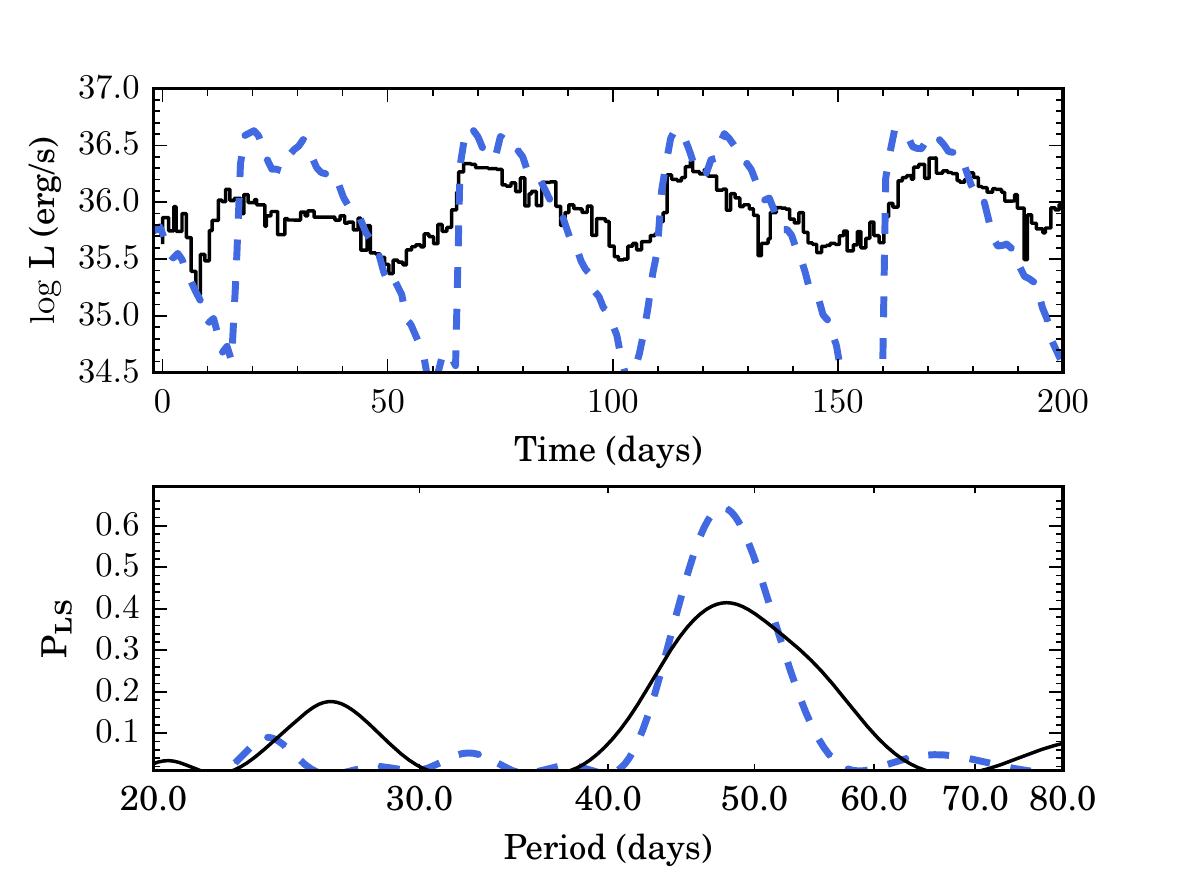}}
\centerline{\includegraphics[width=85mm]{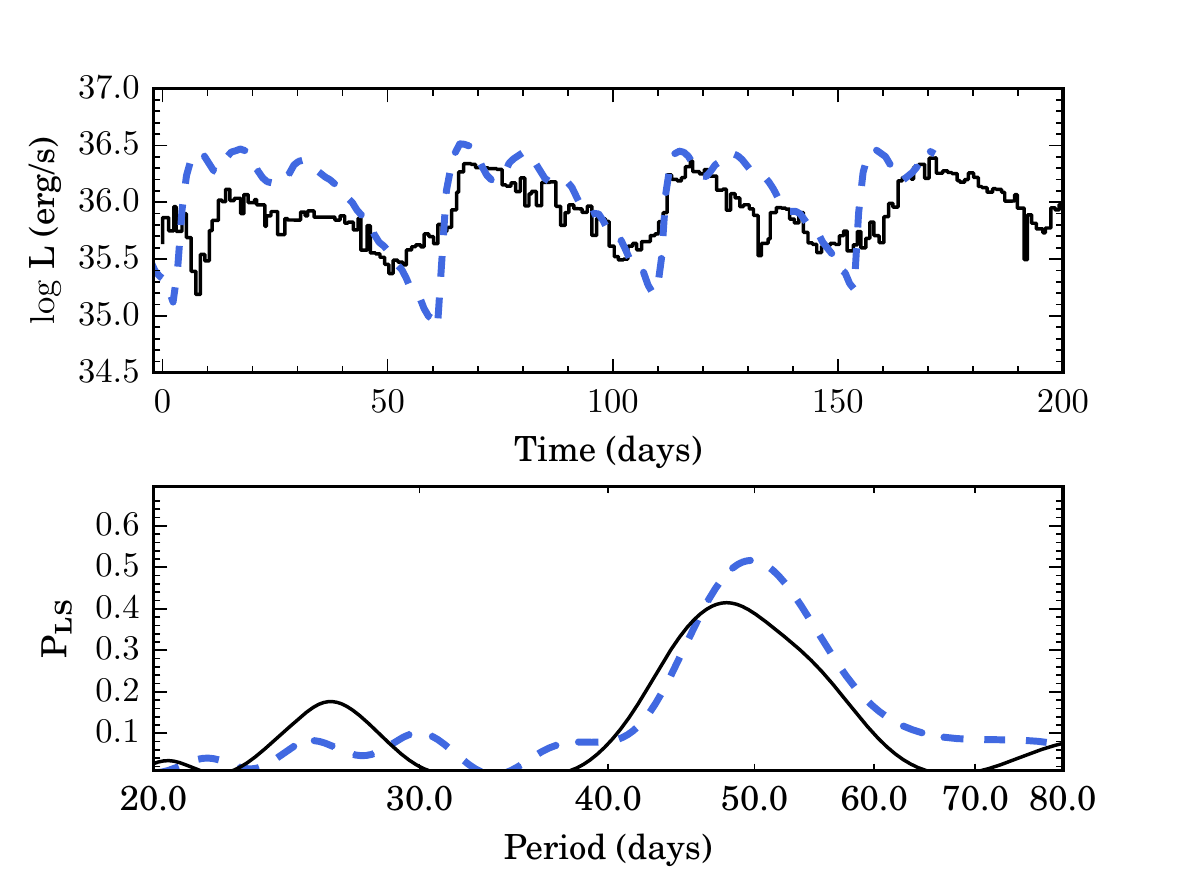}}
\caption{The model 1 for $\alpha_{\rm cold}=0.01$, $\alpha_{\rm hot}=0.05$ (the two upper panels), and model~2 for $\alpha_{\rm cold}=0.01$, $\alpha_{\rm hot}=0.03$ (the two lower panels). For each model, we show the light curve and the corresponding power spectrum in the upper and lower panel, respectively. The solid curves show the observed light curve, given by the sum of one-day average ASM and BAT fluxes, with the time of zero corresponding to MJD 53755. The dashed curves show the theoretical light curves and their power spectra. }
\label{models}
\end{figure}

We have selected time span from 20 January 2006 to 16 August 2006 for the ASM and the BAT light curves, and we summed the fluxes from them as shown in Fig.\ \ref{flux_asm_bat}. This light curve shows relatively regular outbursts with high amplitude. As we found in Section \ref{analysis}, the quasi-period of this part of the data is $P \simeq 45$\,d. We have calculated a large numbers of models, comparing them to the data. 

We show here two models, 1 and 2, in Fig.\ \ref{models}. For each model, we show the light curve and the corresponding power spectrum in the upper and lower panel, respectively. We see that our model 1, with $\alpha_{\rm cold}=0.01$, $\alpha_{\rm hot}=0.05$, reproduces correctly the observed outburst period, but it significantly overestimates its amplitude. Then, our model 2, with $\alpha_{\rm cold}=0.01$ and $\alpha_{\rm hot}=0.03$, approximately reproduces both the period and its amplitude. In this model, the instability begins at $\ga 0.5 R_{\rm out}$.

We find that the outburst amplitude decreases with the increasing $\alpha_{\rm cold} / \alpha_{\rm hot}$. Then, $\alpha_{\rm cold}$ determines the interval between outbursts and $\alpha_{\rm hot}$, the time scale of a single outburst. Both time scales decrease with the increasing value of $\alpha$. We assume here is the constancy of the average $\dot M$, which is justified by the constancy of the flux averaged over an outburst in the chosen data interval. However, changes of the $\dot M$ averaged on time scales longer than that of the outburst will lead to changes of the quasi-period and the amplitude of an individual outburst, as seen in the data. Specifically, for our chosen parameters, an increase of $\dot M$ above $1.5\times 10^{-9}\msun\,{\rm yr}^{-1}$ reduces the outburst amplitude and the duration of the minimum, while it increases the duration of the maximum. A decrease of $\dot M$ leads to the opposite behaviour. The instability disappears at $\dot M\ga 5\times 10^{-9}\msun\,{\rm yr}^{-1}$. Thus, our model can explain the overall properties of the entire data set.

We note that \citet{shih2005} proposed that the observed quasi-periodicity in 4U~1636--536 is explained by an instability due to the irradiation of the disc by X-rays. However, they did not perform any calculations to support their proposal. Our model does not invoke irradiation. We note that irradiation, due to the additional heating imposed on the disc by the X-rays, has a stabilizing effect. Therefore, it may still change the quantitative details of the disc instability model \citep{lasota2015}. As it was shown, e.g., by \citet{dubus2001}, somewhat longer decay-phase and recurrence times are obtained if irradiation is taken into account and the inner disc is allowed to be truncated between the outbursts. In our calculations, we do not take the disc truncation and irradiation into account.

The different values of $\alpha$ in the cold and hot states that we found here to be necessary, are in general agreement with these suggested by observations of dwarf novae and soft X-ray transients.  As it was found by \citet{hirose2014}, an intermittent thermal convection on the upper stable branch of the disc stability curve may strengthen the magnetic turbulence and significantly increase the value of $\alpha$. Nevertheless, the values of $\alpha$ found here are relatively low, in the range of 0.01--0.03 for our best model, and the increase of its value on the hot branch is relatively modest. This means that the values of the Maxwell and Reynolds stresses in the cold and hot disc states are relatively similar, as it was also found in some shearing-box simulations (see \citealt{king2007} and references therein). Thus, at least for the neutron star binary studied here, the poloidal magnetic field required to enhance the angular momentum transport in the ionised disc cannot be very large, and the local MRI turbulence is sufficient to account for its behaviour. In particular, the results obtained here imply the value of $\beta$, the gas to magnetic pressure ratio, to be on the order of $\sim 1/(2\alpha)$ \citep{yuan2014}, which corresponds to $\beta \sim 15$--50. This gives tighter constraints than those resulting from other numerical simulations (see \citealt{yuan2014} for a review).

\section{Conclusions}
\label{conclusions}

We have analysed all the currently available X-ray data of LMXB 4U~1636--536 from the \xte/ASM, \swift/BAT and MAXI detectors, spanning together the energy range of 1.3--50 keV, and observing the source from 1996 to 2014. We have confirmed the previous finding of an appearance of the quasi-periodicity in the X-rays from the source in 2004 \citep{shih2005} and found it continued until 2006. The average period during that epoch is $\simeq$45 d. However, the quasi-period drifts (increasing up to $\sim$70~d or decreasing up to $\sim$37~d) or disappears altogether, and its amplitude changes after 2006. The period of $\simeq$45-d appears again during 2010--2011. 

We then have considered a model of the hydrogen-ionisation disc instability. Using the formalism proposed by \citet{smak1984} to explain outbursts in cataclysmic variables, we have found that it can be successfully used for this accreting neutron-star. In order to reproduce a part of the X-ray light curve of 4U~1636--536 with a stable quasi-periodicity, we have numerically determined the vertical structure and calculated the radial time-dependent evolution of the accretion disc. We have estimated the average accretion rate based on the source distance, the X-ray light curve, and the accretion efficiency for the assumed mass and radius of the neutron star. The main free parameters of our model are then the values of the viscosity parameter, $\alpha$.  

We have found that, in order to reproduce the observed light curve given the source parameters, we need to use two different values of $\alpha$ for the cold (with mostly neutral hydrogen) and hot (with ionized hydrogen) phases, which confirms a number of previous findings, see, e.g., \citet{lasota2001}, \citet{done2007}. Among the large number of the calculated models, the model with relatively low viscosity parameters, $\alpha_{\mathrm{cold}}=0.01$, $\alpha_{\mathrm{hot}}= 0.03$ best reproduces both the amplitude and the period of the used part of the X-ray light curve. 

\begin{acknowledgements}
We would like to thank Agata R\'o\.za\'nska, Bo\.zena Czerny and Monika Mo\'scibrodzka for helpful discussions. This work was partially supported by the POMOST/2012-6/11 Program of Foundation for Polish Science co-financed by the European Union within the European Regional Development Fund (MW and DGR) and by Polish National Science Centre grant DEC-2011/03/D/ST9/00656 (AS). AJ was partially supported by the Polish National Science Center grant 2012/05/E/ST9/03914, and AAZ was supported in part by the Polish NCN grants 2012/04/M/ST9/00780 and 2013/10/M/ST9/00729.
Calculations were performed on PIRXGW computer cluster founded by Foundation for Polish Science within FOCUS program.
\end{acknowledgements}

\bibliographystyle{aa}
\bibliography{4u1636_final}

\end{document}